\documentclass[conference]{IEEEtran}
\IEEEoverridecommandlockouts
\usepackage{cite}
\usepackage{amsmath,amssymb,amsfonts}
\usepackage{algorithmic}
\usepackage{graphicx}
\usepackage{textcomp}
\usepackage{xcolor}
\def\BibTeX{{\rm B\kern-.05em{\sc i\kern-.025em b}\kern-.08em
    T\kern-.1667em\lower.7ex\hbox{E}\kern-.125emX}}

\usepackage{makecell}
\usepackage{amsmath}
\usepackage{amsmath, amssymb, amscd, amsthm, amsfonts}
\usepackage{colortbl}
\usepackage{soul}

\usepackage[colorlinks=true,linkcolor=blue,anchorcolor=black,citecolor=red,filecolor=black,menucolor=black,runcolor=black,urlcolor=black]{hyperref}
    
\begin{document}

\title{Machine Learning based Post Event Analysis for Cybersecurity of Cyber-Physical System  \\
%\thanks{This work was supported by the Korea Institute of Energy Technology Evaluation and Planning (KETEP) and the Ministry of Trade, Industry and Energy (MOTIE) of the Republic of Korea under grant No. 20206910100020.}

\author{\IEEEauthorblockN{Kuchan Park, Junho Hong, Wencong Su}
\IEEEauthorblockA{\textit{Department of Electrical and Computer Engineering} \\
\textit{University of Michigan-Dearborn}\\
Dearborn, USA \\
kuchan@umich.edu, jhwr@umich.edu, wencong@umich.edu} 
\and
%\IEEEauthorblockN{Wencong Su}
%\IEEEauthorblockA{\textit{Electrical and Computer Engineering} \\
%\textit{University of Michigan-Dearborn}\\
%Dearborn, USA \\
%wencong@umich.edu}
\and
\IEEEauthorblockN{
HyoJong Lee}
\IEEEauthorblockA{\textit{New Technology} \\
\textit{DTE Energy}\\
Detroit, MI, USA \\
hyojong.lee@dteenergy.com}
}

%\thanks{\hl{need to write}This work was supported by the Korea Institute of Energy Technology Evaluation and Planning (KETEP) and the Ministry of Trade, Industry and Energy (MOTIE) of the Republic of Korea under Grant No. 20206910100020.}
%\thanks{Corresponding author: Yong-Hwa Kim (e-mail: yongkim@ut.ac.kr).}

%\thanks{K. Park, J. Hong, and W. Su are with the Department of Electrical and Computer Engineering, University of Michigan -- Dearborn, Dearborn, MI, 48128 USA. (emails: kuchan, jhwr, wencong@umich.edu)}
%\thanks{H. Lee is with the New Technology Group, DTE Energy, Detroit, MI, 48226 USA. (email: hyojong.lee@dteenergy.com)}
%\thanks{\hl{need to write}Y.-H. Kim is with Korea National University of Transportation, Uiwang-si, Gyeonggi-do 16106, Republic of Korea. (yongkim@mju.ac.kr)}
}

%\author{\IEEEauthorblockN{Kuchan Park, \textit{Graduate Student Member, IEEE}, Wencong Su, \textit{Senior Member, IEEE}, Junho Hong, \textit{Senior Member, IEEE}}, HyoJong Lee, \textit{Senior Member, IEEE}}

%\author{\IEEEauthorblockN{Aydin Zaboli, Junho Hong}
%\IEEEauthorblockA{\textit{Dep. of Elec. and Comp. Eng.} \\
%\textit{University of Michigan-Dearborn}\\
%Dearborn, MI \\
%{{azaboli, jhwr}}@umich.edu}
%\and
%\IEEEauthorblockN{Junho Hong}
%\IEEEauthorblockA{\textit{ECE} \\
%\textit{University of Michigan-Dearborn}\\
%Dearborn, MI \\
%{{jhwr}}@umich.edu}
%\and
%\IEEEauthorblockN{Vo-Nguyen Tuyet-Doan}
%\IEEEauthorblockA{\textit{Dep. of Electronic Eng.} \\
%\textit{Myongji University}\\
%Yongin-si, South Korea \\
%{{tuyetdoan201096@gmail.com}}
%}
%\and
%\IEEEauthorblockN{Yong-Hwa Kim}
%\textit{Korea National University}\\ \textit{of Transportation} \\
%Gyeonggi-do, South Korea \\
%{{yongkim@mju.ac.kr}}
%}
%\and
%\and

\maketitle

\begin{abstract}
As Information and Communication Technology (ICT) equipment continues to be integrated into power systems, issues related to cybersecurity are increasingly emerging. Particularly noteworthy is the transition to digital substations, which is shifting operations from traditional hardwired-based systems to communication-based Supervisory Control and Data Acquisition (SCADA) system operations. These changes in the power system have increased the vulnerability of the system to cyber-attacks and emphasized its importance. This paper proposes a machine learning (ML) based post event analysis of the power system in order to respond to these cybersecurity issues. An artificial neural network (ANN) and other ML models are trained using transient fault measurements and cyber-attack data on substations. The trained models can successfully distinguish between power system faults and cyber-attacks. Furthermore, the results of the proposed ML-based methods can also identify 10 different fault types and the location where the event occurred. 
\end{abstract}


\section{Introduction}
In the power system architecture, the transmission network plays a pivotal role in efficient electricity delivery with minimized losses. Protecting this transmission infrastructure is thus crucial for maintaining the overall stability and reliability of the power system~\cite{10348488}. Faults in the power system can arise from various factors, such as heavy snowfall, lightning, earthquakes, or animal interference. Inadequate or delayed responses to these incidents can lead to extended power outages, irreversible equipment damage, and significant economic repercussions. The implications of such damage extend beyond the transmission system, impacting both power generation and distribution sectors. With the evolution of the power system towards a more distributed framework as opposed to the traditional centralized model, the system dynamics have become increasingly complex. This complexity introduces heightened vulnerabilities and necessitates rapid response mechanisms during emergencies. The deployment of AI-based technologies for fault and cyber-attack detection, enabled by recent advancements in big data storage and processing capabilities, is essential to address these challenges effectively.

There are several recent related works about this field. The works of~\cite{9395448} developed artificial intelligence based intrusion detection system. Sampled value (SV) message of IEC61850 communication protocol and penetration of renewable energy were considered. A cyber-attack detection technology for Transmission Protective Relays based on deep learning techniques has been proposed by~\cite{9270592}. The authors of~\cite{en15155539} presents a method that involves preprocessing power system data to convert it into image-like formats, enabling the use of Convolutional Neural Networks (CNN) to detect faults and cyber attacks. This innovative approach leverages the powerful image processing capabilities of CNNs to analyze and interpret complex power system data. However, such studies have faced challenges in accurately detecting the types of faults and determining their locations within power systems. Additionally, being primarily focused on relays, these research efforts are limited in scope. As a result, applying these methods to the broader context of the entire power system presents difficulties, highlighting the need for more comprehensive and adaptable solutions.

Therefore, this paper proposes a fault and cyber attack detection methods based on machine learning algorithm. The IEEE 14 bus system has been used to generate a power system fault and cyber attack (e.g., abnormal data injection attack) data for training and validation. The generated dataset is used to train ANN and other ML models. The proposed ML-based post-fault analysis framework can help operators initiate the post event study more efficiently than traditional methods. Three-phase voltage and current values were measured after faults in the power system to generate data to train machine learning models. The proposed method is capable of identifying ten different types of faults and can also determine the location of the fault along the line.  In the scenario of a cyber-attack, the technique of False Data Injection (FDI) was employed~\cite{9401940}. For instance, it was hypothesized that a cyber attacker successfully accessed a digital substation and compromised the merging unit (MU). Subsequently, the attacker injected abnormal sampled value messages into the substation network. This led to a situation where the protective intelligent electronic device (PIED) misinterpreted the injected sampled value (SV) messages as legitimate fault current and voltage values, resulting in an erroneous tripping of the circuit breaker. Additionally, to assess whether the proposed method remains effective under N-1 contingency conditions, it was applied to a modified topology of the power system (classifying faults and cyber-attacks with one transmission line out of service). The methodology was tested and validated using the IEEE 14 bus system with Opal-RT real-time simulator. The test results demonstrated that the proposed approach successfully distinguished between cyber-attacks and conventional power system faults, including their locations, even while considering N-1 contingency scenarios. Key contributions of this paper are as follows:
\begin{itemize}
    \item The traditional methods for diagnosing faults in power systems faced challenges in differentiating between cyber-attacks and regular faults. This difficulty arises because a fault data injection attack can accurately replicate actual fault current and voltage. However, the proposed method leverages machine learning to compare voltage, current, angle, and frequency data from the attacked substation and its neighboring substations, thereby determining whether an event is a genuine fault or a cyber-attack.
    \item If the event is identified as an actual fault, the method considers a total of ten fault types, enabling it to determine the specific type of fault that caused the circuit breaker to trip, as well as the location of the fault.
    \item Although this paper focuses solely on N-1 contingency scenarios, future applications could extend to N-2 or even N-M contingencies, encompassing larger-scale disruptions caused by cyber-attacks. Additionally, while this study validates the proposed algorithm using the IEEE 14 bus system, the approach is scalable and can be applied to larger systems, such as the IEEE 39 or IEEE 118 bus systems.
    \end{itemize}

%This research employs a stacked LSTM autoencoder (LSTM-SAE) topology to investigate the load forecasting for two residences regarding various PV, BESS, and EV arrangements. In this paper, a sequence to sequence design-based on LSTM is proposed to anticipate the alterations in electric current patterns based on the users' behavior. Furthermore, the trained data is based on 1-minute intervals that can provide better accuracy for the prediction process. It aims the utilities to acquire smart meter data promptly, so they can accurately analyze the load patterns of their customers, which could be helpful for both utilities and customers to check the possible anomaly in load behavior simultaneously. A recurrent neural network-based LSTM-SAE is presented to forecast how different DER configurations will affect the complication of a BTM system. This developed LSTM model's performance assessment reveals great effectiveness and precision.

The remaining part of this paper is organized as follows: Section II illustrates a proposed post event analysis. The validations are mentioned in Section III. Finally, this paper is concluded in Section IV.

\section{PROPOSED POST EVENT ANALYSIS}

\begin{figure}[!t]
\centerline{\includegraphics[width=0.65\columnwidth]{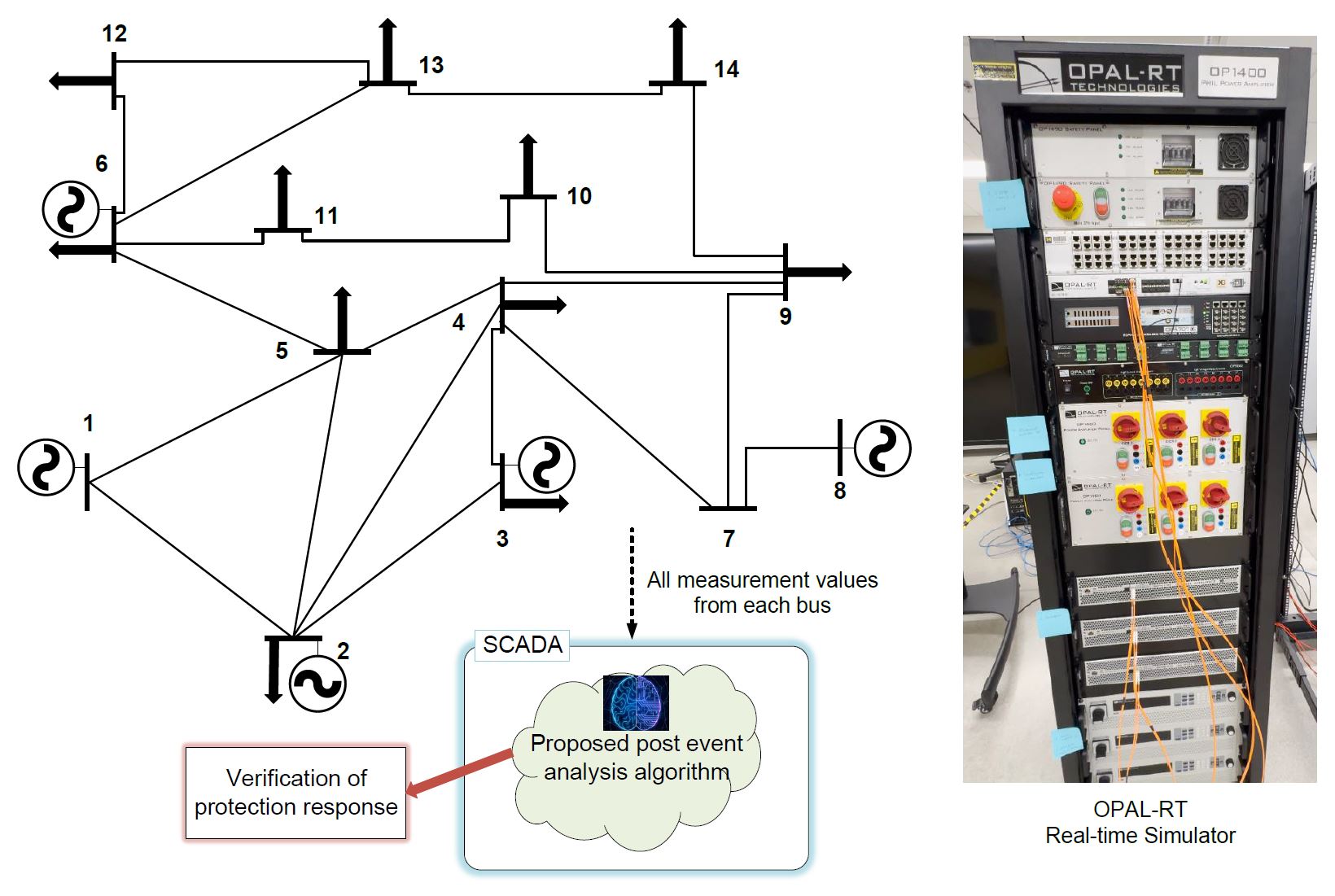}}
\caption{An example of post event analysis applied to an IEEE 14 bus system.}
\label{fig:ieee14}
\end{figure}

As communication and scientific technologies advance, the patterns of cyber-attacks are becoming more diverse. For example, when an attack targets the Sampled Value (SV) messages of a Merging Unit (MU), it is extremely challenging for the Protective Intelligent Electronic Device (PIED) that subscribes to these SV messages to distinguish between a cyber-attack and actual fault currents and voltages. Therefore, this paper proposes an algorithm using MLs to discern the true nature of such events. Fig. \ref{fig:ieee14} in the paper displays the single-line diagram of the IEEE 14 bus power system used in this study~\cite{9633750}.

Typically, when a fault occurs in a power system, the PIED is expected to detect the fault and issue a trip signal to the circuit breaker within 0.1 to 3 cycles. The transient waveforms of these faults can be crucial in determining whether an event is a cyber-attack or a regular fault. For instance, consider a hypothetical fault occurring between buses 4 and 5 in Fig. \ref{fig:ieee14}. Following the fault, all neighboring buses (such as buses 1, 2, 3, 6, 7, and 9) would experience similar transient fault waveforms. However, in the case of a cyber-attack leading to a trip event, only the targeted bus would show the fault transient waveforms, while the neighboring buses would not. This paper proposes a technique that utilizes this power system domain knowledge in combination with machine learning to analyze such events.

\subsection{Proposed Post Event Analysis}

\begin{figure}[!t]
\centerline{\includegraphics[width=0.6\columnwidth]{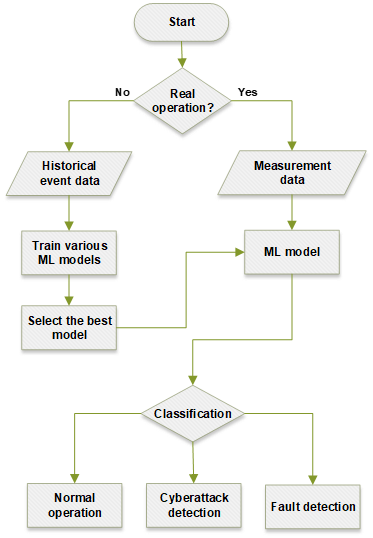}}
\caption{The flowchart of proposed post event analysis.}
\label{fig:flow}
\end{figure}

Fig. \ref{fig:flow} in this paper presents a flowchart of the proposed methodology. The overall approach is divided into two parts. The left side of the flowchart illustrates the process of training ML models and selecting the most effective one. The right side demonstrates the application of the trained ML models for real-time event analysis. Historical event data and fault simulation data were used for training the ML models. Separate simulations were conducted to generate data for cyber-attacks, specifically targeting the Sampled Value (SV) messages of Merging Units (MUs).

Four different ML models were employed, each optimized through hyperparameter tuning. To identify the optimal ML model, a comparative analysis of the four models was conducted, with the selection based on the best performance. As new ML models emerge, they can be evaluated using this methodology to determine if they offer improved results compared to the currently selected optimal model.

Once the optimal ML model was chosen, it was applied in real-time applications. The generated data (after a circuit breaker trip) was transmitted in real-time to the ML model. Then the ML model demonstrated its capability to discern among ten different types of faults, the location of faults, cyber-attacks, and steady-state operations.

\subsection{Machine Learning Algorithm Analysis for Fault Type and Location Classification}

As explained, the proposed framework trains several machine learning models and selects the one that shows the best performance. Therefore, several representative ML algorithms were utilized in the proposed post event analysis.

The first is Decision Tree (DT). A decision tree model classifies samples based on characteristics~\cite{10015734}. It is called a decision tree because its shape resembles the branches of a tree. It is a process of finding the correct answer class by asking questions with the values of the attributes. The correct answer to a question or terminal is called a node, and the line connecting the nodes is called an edge. DT mainly uses gini index or entropy as a loss function. In the proposed method, the following gini index is used.

\begin{equation} \label{E1}
    gini(D)=1-\sum_{j=1}^{n}p_{j}^{2},
 \end{equation}

 where $D$ denotes the number of data, $n$ represents the number of classes, $p$ means the ratio of the data belonging to class $j$.

The second is Support Vector Machine (SVM). SVM is a method that finds the best boundary, i.e. the hyperplane, to classify classes~\cite{10329926}. A gaussian SVM is applied, which can classify data with non-linear features.

 \begin{equation} \label{E2}
    \underset{\alpha }{max}\sum_{i=1}^{m}\alpha _{i}-\frac{1}{2}\sum_{i=1}^{m}\; \sum_{j=1}^{m}\alpha _{i}\alpha _{j}y_{i}y_{j}K(x_{i}\cdot x_{j}).
 \end{equation}
 \begin{equation*} \label{E2-1}
   subject\: to\: \alpha _{i}\geq 0,\: \sum_{i=1}^{m}\alpha _{i}y_{i}=0.
 \end{equation*}

 Eq. 2 is the objective function of a typical SVM. $\alpha$ denotes the Lagrange multiplier, $m$ represent the number of data, $(x,y)$ is the data point, and $K$ means the Gaussian kernel.

 A K-nearest neighbors (KNN) which predicts outputs based on K nearest neighbors is applied as a third model. KNN uses a distance metric to find which input data is most similar to the trained data. Minkowski distance is used as the distance function used for KNN.

 \begin{equation} \label{E3}
   dist(x,y)=\left ( \sum_{i=1}^{n}\left| x_{i}-y_{i}\right|^{p} \right )^{1/p},
 \end{equation}

where $x,y$ represent data points, $n$ denotes the dimension, and $p$ means the order of the norm.

Finally, an artificial neural network (ANN) is trained. An ANN is a type of computing system that organizes patterns to mimic the behavior of neurons in the human brain. Based on input information, it determines which class that input belongs to. Classification is implemented by comparing the values of multiple output nodes and selecting the greater one. The loss function evaluates how much the error varies based on the weights and often uses a cross-entropy function. The probability value is obtained by the Softmax calculation.

\begin{table}[]
\centering
\caption{System status information}
\label{T1}
\begin{tabular}{|c|c|}
\hline
\textbf{Class index} & \textbf{Description} \\ \hline
0                    & Normal operation     \\ \hline
1                    & A-gnd fault          \\ \hline
2                    & B-gnd fault          \\ \hline
3                    & C-gnd fault          \\ \hline
4                    & AB-gnd fault         \\ \hline
5                    & BC-gnd fault         \\ \hline
6                    & AC-gnd fault         \\ \hline
7                    & A-B fault            \\ \hline
8                    & B-C fault            \\ \hline
9                    & A-C fault            \\ \hline
10                   & ABC-gnd fault        \\ \hline
11                   & Cyber attack          \\ \hline
\end{tabular}
\end{table}

\subsection{Cyber attack Implementation}

\begin{figure*}[t]
\centering
\includegraphics[width=1.6\columnwidth]{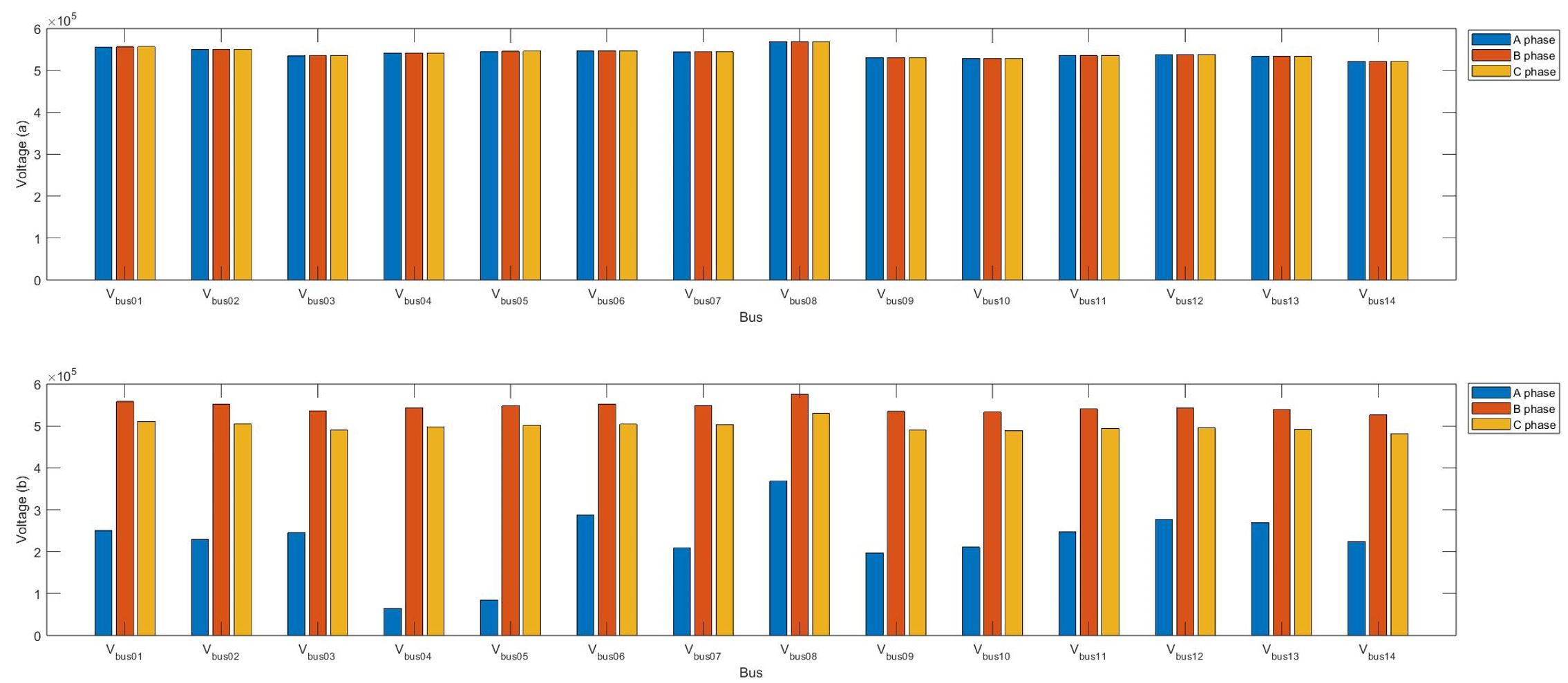}
\caption{\centering {An example of three phase bus voltages: (a) without fault and (b) fault at line 4-5.}
%ISO TARA. 
}
\label{fig:bus_voltage}
\end{figure*}

\begin{figure}[t]
\centering
\includegraphics[width= .4\textwidth, height = 1.4 in]{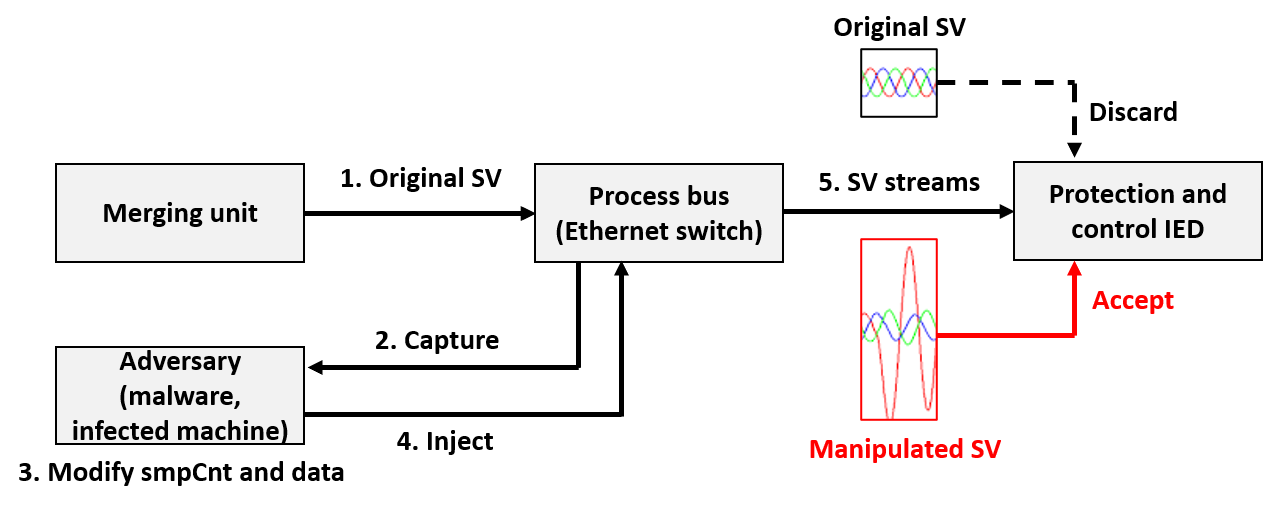}
\caption{An example of spoofing attacks for SV messages \cite{9361308}.}
\label{fig:SpoofingSV}
\end{figure}

In this paper, as shown in Table \ref{T1}, a total of 12 classes were established. Class 0 represents normal operation, where there are no faults or cyber-attacks. Classes 1 to 10 correspond to ten distinct types of faults. Class 11 is designated for cyber-attacks. To make it challenging for the Protective Intelligent Electronic Device (PIED) to differentiate between actual power system faults and cyber-attacks, a method using real fault current and voltage data was implemented for Sampled Value (SV) fault data injection attacks. However, as mentioned earlier, since these cyber-attacks target a single substation, the neighboring buses do not exhibit fault transients, which is a key distinguishing factor in the proposed method. For the cyber attack, a reply and false data injection attacks were executed as shown in Fig. \ref{fig:SpoofingSV}. A replay attack can be initiated by playing back older SV packets that contain fault currents and voltages, which are critical information to pass on. In order to achieve the replay attack, attackers need to gain access to the monitoring port of process bus Ethernet switch, and capture the critical status of SV messages. The expected impact of a successful attack is to open the circuit breakers by triggering the protection functions of the SV subscriber (PIED).
% In this study, measured parameters from all buses in the transmission system are sent to the SCADA system. The communication between the SCADA and the subsystems on each bus is usually performed using the DNP3 protocol. Therefore, there are two possibilities for these measurements to be corrupted by cyberattacks. First, the measuring devices on each bus are directly attacked and measure the wrong information. The second possibility is that the attacker tampered with the DNP3 communication packets, causing the SCADA to make incorrect decisions.

\section{CASE STUDIES}
\subsection{System Configuration}
As illustrated in Fig. \ref{fig:flow}, to select the optimal machine learning (ML) model and apply it to a real-time system, a hardware-in-the-loop (HIL) testbed was constructed using OPAL-RT, as shown in Fig. \ref{fig:ieee14}. For data generation, each of the 20 lines was assigned four different fault locations (20\%, 40\%, 60\%, and 80\%) and four different fault impedances (0.001, 0.05, 0.1, and 0.15) were used, thereby simulating a total of ten different fault types. Cyber-attack scenarios were generated for each case, completing a dataset of 6,400 samples. In total, 202 classes were utilized to distinguish between types of faults, fault locations, and cyber-attacks. Fig. \ref{fig:bus_voltage} shows an example of three phase voltages at each bus with and without fault at line 4-5.

In the real-time tests, the trip signal from the Protective Intelligent Electronic Device (PIED) was designated as the starting point for the post event analysis proposed in this paper. The phasor measurement unit (PMU) data corresponding to the time of the trip signal were used as input data (30 data per second). Thus, the input data for the ML model in real-time operations include voltage (V), current (I), angle, and frequency values from all buses where PMUs are installed.

\subsection{Single Fault Post Event Analysis Result}
Table \ref{T2} displays the performance metrics of four different machine learning (ML) models for single fault post event analysis result. The data generation and training/validation of the ML models were conducted by considering only a single fault scenario. This approach focuses on a specific type of fault, allowing for a detailed and concentrated analysis of the ML model's performance in identifying and responding to that particular fault condition in the power system. In the process of selecting the optimal machine learning (ML) model, the dataset was partitioned such that 90\% was used for training and the remaining 10\% for validation. This approach ensures a comprehensive evaluation of the ML models, balancing the need for a robust training dataset with a sufficiently large validation set to assess the model's performance and generalization capabilities accurately.

While Decision Trees (DT) and Support Vector Machines (SVM) share the same accuracy, SVM shows relatively higher precision, recall, and F1-score. Similarly, the precision, recall, and F1-score for K-Nearest Neighbors (KNN) are lower compared to SVM. This is because DT and KNN mistakenly classify normal operation data as cyber attack situations. In contrast, SVM achieves better differentiation between normal operations and cyber attacks, hence the higher performance metrics. The Artificial Neural Network (ANN) model outperforms the others, with all metrics reaching 100\%. Consequently, with its superior performance, the ANN was selected for application in real-time operations.

\begin{table}[]
\centering
\caption{Model performance of post event scenario}
\label{T2}
\begin{tabular}{c|c|c|c|c|}
\cline{2-5}
 & \textbf{Accuracy(\%)} & \textbf{Precision(\%)} & \textbf{Recall(\%)} & \textbf{F1-score(\%)} \\ \hline
\multicolumn{1}{|c|}{\textbf{DT}}  & 97.8 & 90.49 & 90.10 & 90.30 \\ \hline
\multicolumn{1}{|c|}{\textbf{SVM}} & 97.8 & 98.32 & 96.35 & 97.32 \\ \hline
\multicolumn{1}{|c|}{\textbf{KNN}} & 98.8 & 91.46 & 91.67 & 91.56 \\ \hline
\multicolumn{1}{|c|}{\textbf{ANN}} & 100  & 100   & 100   & 100   \\ \hline
\end{tabular}
\end{table}

\subsection{N-1 Contingency Event (with Single Fault) Analysis Result}
Table \ref{T3} presents the training results using data collected from N-1 contingency scenarios. 
For example, the paper examined whether the proposed method could still be effectively applied when the topology of the power system changed due to the tripping of a single line. This assessment is crucial for understanding the robustness and adaptability of the proposed solution in dynamic operational environments, where changes in system topology can have significant impacts on fault detection and response mechanisms. Among the four models, the Decision Tree (DT) exhibits the lowest performance with an accuracy of 91.3\%, and it also scores the lowest across the other three metrics. The K-Nearest Neighbors (KNN) model shows slightly better metrics compared to DT but significantly lags behind Support Vector Machines (SVM) and Artificial Neural Networks (ANN). SVM demonstrates respectable results with 98\% accuracy, 99.67\% precision, 95.83\% recall, and 97.71\% F1-score. However, ANN outperforms all with 100\% across all metrics, establishing it as the model with the best performance in the N-1 contingency context.

\begin{table}[]
\centering
\caption{Model performance of N-1 contingency scenario}
\label{T3}
\begin{tabular}{c|c|c|c|c|}
\cline{2-5}
 & \textbf{Accuracy(\%)} & \textbf{Precision(\%)} & \textbf{Recall(\%)} & \textbf{F1-score(\%)} \\ \hline
\multicolumn{1}{|c|}{\textbf{DT}}  & 91.3 & 85.06 & 87.63 & 86.33 \\ \hline
\multicolumn{1}{|c|}{\textbf{SVM}} & 98   & 99.67 & 95.83 & 97.71 \\ \hline
\multicolumn{1}{|c|}{\textbf{KNN}} & 92.6 & 92.12 & 91.88 & 92.00 \\ \hline
\multicolumn{1}{|c|}{\textbf{ANN}} & 100  & 100   & 100   & 100   \\ \hline
\end{tabular}
\end{table}

\subsection{Simultaneous Event (Two Faults) Analysis Result}
In scenarios such as heavy snowfall, which may cause the collapse of aging transmission facilities, or widespread natural disasters (e.g., wildfires and hurricanes) can occur simultaneously at multiple locations. Such emergency situations can drastically increase the number of potential scenarios, potentially making the system more vulnerable to cyber attacks. Moreover, if well-trained cyber attackers simultaneously target two substations, it becomes crucial to test whether the proposed algorithm can accurately differentiate between cyber attacks and actual faults.

To address these challenges and accurately diagnose faults while filtering out cyber attacks in such situations, a simulation of simultaneous event analysis was conducted. This approach is essential to ensure that the power system's protective mechanisms remain effective and resilient, even under complex and simultaneous fault conditions combined with potential cyber threats.

\begin{table}[]
\centering
\caption{Model performance of Simultaneous Event scenario}
\label{T5}
\begin{tabular}{c|c|c|c|c|}
\cline{2-5}
 & \textbf{Accuracy(\%)} & \textbf{Precision(\%)} & \textbf{Recall(\%)} & \textbf{F1-score(\%)} \\ \hline
\multicolumn{1}{|c|}{\textbf{DT}}  & 95.00 & 91.04 & 93.60 & 92.30 \\ \hline
\multicolumn{1}{|c|}{\textbf{SVM}} & 99.00 & 98.84 & 98.78 & 98.81 \\ \hline
\multicolumn{1}{|c|}{\textbf{KNN}} & 91.50 & 83.29 & 89.02 & 86.06 \\ \hline
\multicolumn{1}{|c|}{\textbf{ANN}} & 99.30 & 99.29 & 98.88 & 99.09 \\ \hline
\end{tabular}
\end{table}

Table \ref{T5} shows the performance of the machine learning models for the simultaneous event scenario. In this case, the worst performing model is KNN and the best performing model is ANN. SVM performs as well as ANN, but the overall values is slightly higher for ANN. DT performs well compared to the previous scenario, but performs poorly compared to SVM and ANN.

\section{Conclusion}
 As the power system becomes more decentralized and ICT system evolves, the significance of cybersecurity is increasingly emphasized. In line with this trend, this paper proposes a ML-based post event analysis method to detect cyber-attacks and accurately diagnose faults. Historical fault data, cyber-attack data, and simulation data from HIL testbed were collected and used to train various ML models. Among these models, the one demonstrating the best performance was selected for application in real-time operations.

Once an event occurs (CB trip), PMU readings from substations at each bus of the transmission system are received via the SCADA system. The voltage, current, angle, and frequency values from each phase over three cycles are input into the selected ML model to distinguish between cyber-attacks and faults (identifying its fault type and location). The proposed method was validated using the IEEE 14 bus system. Four ML models, e.g., Decision Tree (DT), Support Vector Machine (SVM), K-Nearest Neighbors (KNN), and Artificial Neural Network (ANN), were considered, and the best-performing model was applied to real-time operations. The algorithm was tested and validated by various scenarios, including single fault, N-1 contingency, and simultaneous event scenarios.

For future work, data from larger systems (such as the IEEE 39 or IEEE 118 bus systems) will be used/collected, and more diverse ML models will be trained to apply the proposed method. Additionally, the method will be validated in scenarios with a higher penetration of renewable energy sources. For verification in real-time scenarios, a Hardware in the Loop Simulation (HILs) testbed is being constructed using PMU, Programmable Logic Controller (PLC), and a real-time simulator. The proposed algorithm will be verified in a real-time environment through the testbed.

%\begin{thebibliography}{00}
\bibliographystyle{IEEEtran}
\bibliography{IEEEabrv,ref}

\begin{thebibliography}{1}
\providecommand{\url}[1]{#1}
\csname url@samestyle\endcsname
\providecommand{\newblock}{\relax}
\providecommand{\bibinfo}[2]{#2}
\providecommand{\BIBentrySTDinterwordspacing}{\spaceskip=0pt\relax}
\providecommand{\BIBentryALTinterwordstretchfactor}{4}
\providecommand{\BIBentryALTinterwordspacing}{\spaceskip=\fontdimen2\font plus
\BIBentryALTinterwordstretchfactor\fontdimen3\font minus \fontdimen4\font\relax}
\providecommand{\BIBforeignlanguage}[2]{{%
\expandafter\ifx\csname l@#1\endcsname\relax
\typeout{** WARNING: IEEEtran.bst: No hyphenation pattern has been}%
\typeout{** loaded for the language `#1'. Using the pattern for}%
\typeout{** the default language instead.}%
\else
\language=\csname l@#1\endcsname
\fi
#2}}
\providecommand{\BIBdecl}{\relax}
\BIBdecl

\bibitem{10348488}
E.~Xu, H.~Wang, and Y.~Zhang, ``A study on transmission line protection using incident current,'' in \emph{2023 IEEE International Conference on Advanced Power System Automation and Protection (APAP)}, 2023, pp. 371--377.

\bibitem{9395448}
T.~S. Ustun, S.~M.~S. Hussain, L.~Yavuz, and A.~Onen, ``Artificial intelligence based intrusion detection system for iec 61850 sampled values under symmetric and asymmetric faults,'' \emph{IEEE Access}, vol.~9, pp. 56\,486--56\,495, 2021.

\bibitem{9270592}
Y.~M. Khaw, A.~Abiri~Jahromi, M.~F.~M. Arani, S.~Sanner, D.~Kundur, and M.~Kassouf, ``A deep learning-based cyberattack detection system for transmission protective relays,'' \emph{IEEE Transactions on Smart Grid}, vol.~12, no.~3, pp. 2554--2565, 2021.

\bibitem{en15155539}
\BIBentryALTinterwordspacing
J.~Hong, Y.-H. Kim, H.~Nhung-Nguyen, J.~Kwon, and H.~Lee, ``Deep-learning based fault events analysis in power systems,'' \emph{Energies}, vol.~15, no.~15, 2022. [Online]. Available: \url{https://www.mdpi.com/1996-1073/15/15/5539}
\BIBentrySTDinterwordspacing

\bibitem{9401940}
S.~Tufail, S.~Batool, and A.~I. Sarwat, ``False data injection impact analysis in ai-based smart grid,'' in \emph{SoutheastCon 2021}, 2021, pp. 01--07.

\bibitem{9633750}
P.~Kumar, B.~Bag, N.~D. Londhe, and A.~Tikariha, ``Classification and analysis of power system faults in ieee-14 bus system using machine learning algorithm,'' in \emph{2021 4th International Conference on Recent Developments in Control, Automation \& Power Engineering (RDCAPE)}, 2021, pp. 122--126.

\bibitem{10015734}
R.~Valiyil and J.~Gangadharan, ``Rolling average-decision tree-based fault detection of neutral point clamped inverters,'' \emph{IEEE Journal of Emerging and Selected Topics in Industrial Electronics}, vol.~4, no.~3, pp. 744--755, 2023.

\bibitem{10329926}
P.~Sun, X.~Liu, M.~Lin, J.~Wang, T.~Jiang, and Y.~Wang, ``Transmission line fault diagnosis method based on improved multiple svm model,'' \emph{IEEE Access}, vol.~11, pp. 133\,825--133\,834, 2023.

\bibitem{9361308}
J.~Hong, R.~Karnati, C.-W. Ten, S.~Lee, and S.~Choi, ``Implementation of secure sampled value ({S}e{S}{V}) messages in substation automation system,'' \emph{IEEE Transactions on Power Delivery}, pp. 405--414, 2021.

\end{thebibliography}

\end{document}